How IT allows E-Participation in Policy-Making Process

Sourav Mukherjee

Senior Database Administrator &

PhD student at University of the Cumberlands

Chicago, United States

Abstract

With the art and practice of government policy-making, public work, and citizen participation, many governments adopt information and communication technologies (ICT) as a vehicle to facilitate their relationship with citizens. This participation process is widely known as E-Participation or "Electronic Participation". This article focuses on different performance indicators and the relevant tools for each level. Despite the growing scientific and pragmatic significance of e-participation, that area still was not able to grow as it was expected. Our diverse set of knowledge and e-participation policies and its implementation is very limited. This is the key reason why e-participation initiatives in practice often fall short of expectations. This study collects the existing perceptions from the various interdisciplinary scientific literature to determine a unifying definition and demonstrates the strong abilities of e-participation and other related components which have great potential in the coming years.

How IT allows E-Participation in Policy-Making Process

In this information-age, led by Internet content, resources, innovation in software and technology and interdependence to multiple other systems in getting the society changed. E-participation, E-Governance, ICT tools, e-empowerment, e-engagement are the interrelated concepts or models help us effectively manage, participate, govern and pursue the public work. Regardless Of the omnipresence of e-Participation programs, endeavors of social media-centered and citizen-guided political discussions are extremely limited. Therefore, there lies a very little chance to control, study and realize the desired debate regarding the traditional e-Participation and a free or natural citizen debate on the social media platforms. Currently, the interest has made a radical shift towards harnessing the informal channels part of the universal e-Participation solution. Still, the unspoken belief of the duality of e-Participation is yet to be studied, investigated and hypothesized. To implement such "duality of e-Participation", requires a robust social software foundation to empower decision makers in getting access to the government entrance relevant resources regarding the continuing citizen debates on the social media platforms.

**Literature Review or Background**

In the last couple of decades, a steady need to consider the innovative application of ICTs have been evolved to play a part that may enable the broader audience to participate to independent debate, the places where the roles are wider, broader and deeper. This understanding has stemmed in several secluded e-democracy guides and research studies. It is crucial to strengthen this work and distinguishes the degree of participation, the technology utilized, the phase in the policy-making process and numerous issues and restrictions, including the possible benefits. E-democracy is widely used in the ICT model to support and encourage democratic decision-making processes. In a few countries, e-democracy is used synonymously used with e-voting. However, the voting process is just a subset of the entire E-democracy process. Voting is not only the only criteria through which the citizens can impact the policy-making or the decision-making process. Rather e-democracy may be further divided into two separate areas such as e-participation and the other is e-voting. E-Voting has been identified to go under many challenges and difficulties, whereas the e-participation has a greater prospect with regards to discussion and dialogue between the government, policymakers and the citizens. In many countries, the election process is a concern to a high intensity of risk, whether it includes tampering with the voting machines, dishonest vote kiosks/booths, or merely hiding the vote arrangement process of supervision and public inspection. Nowadays because of the advancement in software process, hiding malware in the voting device is no way a difficult task. Tampering with the voting results are far easier and will continue to rise over a period such issue. The malware deployed may have an algorithm which can efficiently change the votes from one candidate to the other and that may do its job very easily without being noticed by the others. Hence it requires a malware detector to quickly identify if the voting machine is free from any such scrupulous software. It has been observed that most of the time the security improvements are driven by any kind of breaches and hence it requires policymakers, government agencies, and security professionals to work together to identify a better option to handle such issues. Due to this reason, e-voting is not preferred. On the other hand, when it comes to the e-participation there is a large magnitude of government, corporations using technology and expertise to provide access to policy information. While some governments and study centers have already carried out several surveys in this area and explored that there is no single approach to describe the problem and detailing the outcome. Therefore e-democracy is involved with the use of information and communication technologies (ICT) to connect citizens supporting the independent decision-making practices and improving the representative republic. The primary ICT method is the internet-based approach, including computers used in the home, office or in public locations, mobile phones, and television. The self-governing decision-making processes can be separated into two major groups: one focus on the electoral process, involving e-voting, and the other adopting citizen e-participation in independent policymaking.

## Discussion

e-governance is the product of information and communication technology (ICT) for providing government public services, information exchange, communication operations, incorporation of various systems and services to customers or to other businesses. This is nothing but governance in an electronic environment using electronic media and tools.

Generally, the goals of the e-Governance are:

- Serving the citizens through a better model
- Leading towards transparency and accountability.
- Enhance efficiency within government
- Develop suitable connectivity between business and industry.

Due to the growing use of computers, Internets, smartphones, changes our lifestyle and the way we undertake and execute any work, learn and respond. Around the world, governments are started to realize the benefits of the value of e-governance. If this is designed and implemented appropriately, the e-governance process can greatly benefit the people in terms of delivering high-quality government services, simplifying the government regulations, improving the citizen participation, creating an enabling trust in the government, and so on. For that reason, policymakers and other officials are looking to implement e-Government in countries around the world.

Regarding e-empowerment, e-petitions and e-referenda are two possible ways for the collection of citizen opinions and remarks to control policy. The expansion of online populations of interest, in which certain policy concerns are debated and different proposals are devised all over again based on discussion forums, are also some instances of empowerment online.

One of the most common ways to empower citizens are E-empowering. In this policy-making process, citizens are involved with encouraging active involvement and assisting bottom-up proposals to sway the political agenda. Since the bottom-up viewpoint, citizens are evolving as producers instead just users of the policy.

Therefore, I think that the levels of participation do include enabling, engaging, and empowering.

Stages in the policy-making process are very complicated as it involved multiple parameters as defined in the below picture to know when to engage the citizens,

- Setting Agenda – identify the problem which we need to address.
- Analysis – define the possible challenges and if there any opportunities associated with the agenda.
- Create policy – create a good workable policy document.
- Implementation – create a solid delivery plan.

- Policy monitoring – involves evaluation and review of the policy.

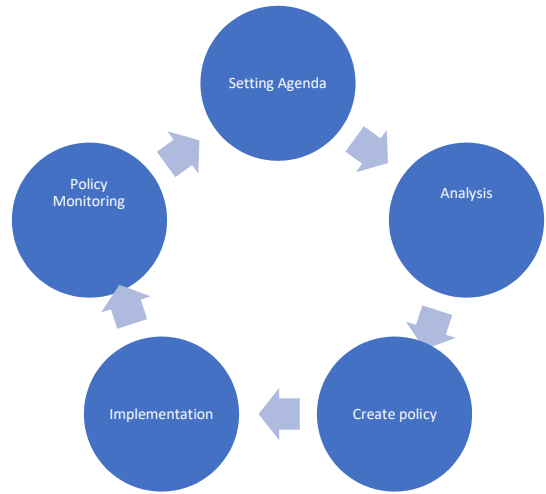

Figure 1. Stages in the Policy-making process

ICT offers the flexibility to policy-makers to reach directly to the users who are using the services. Essentially citizens have more influence on policy matter through discussion before then the policy-making procedure kicks in. Consultation at the phase of an outline policy document involves citizens to have the communication skills to understand the usual legalistic terminology of the document before commenting appropriately. Citizens are required to be well-informed on issues before the above stage is reached and the information must be more accurate, readable and understandable.

ICT can promote empowerment, involvement and getting government processes more effective and understandable by sharing information and communication among the people and organizations within the government. Even governments can improve on the quality and awareness of the services they provide to the citizens and increase the availability of the services and the infrastructure.

This is enabled by e-government applications over the Internet and other communication networks that deliver services and information to people. ICT can bind people and local groups with information and resources outside their geographical limits, inspiring information distribution, knowledge exchange, data, and communication. ICT also encourage the citizens to participate in the democratic process in the form of electronic forums and bulletin boards which facilitate participation in the public conversations. To control political and influential decision-making processes, the organizations in the developing nations like to participate in the information-sharing process that improves governance and collective energy. Even Though people and organizations can efficiently use ICT to enhance their information exchange and communications, robust leadership and executive resources are required to turn information into organized engagement.

**e-participation Tools and Techniques**

Because of the versatility of the nature of the tools and technologies, most of the tools, and the variety of actors in the participation process, it is challenging to identify and realize the connections between citizens and technology. Such element measures how members are involved, and which devices will be possible to help involvement efficiently.

Electronic participation services through ICTs should be positioned at the center of the e-participation process as they form the principle of e-participation. Investigating such tools is deemed a shred of convincing evidence for the significance of the ICT tools in the achievement of e-participation developments.

In the 2000s the European Commission (EC) financed numerous e-participation projects under the e-participation foundation act, the organizing of ICT, its experience, and the newest developments in an online collaboration were the common features of the most popular projects.

Sobaci has established a framework for the applicable ICT tools according to the various e-participation's purposes and the attributes desired to achieve these intentions, also Phang and Kankanhalli suggested another framework that offered five ideas of e-participation and the finest ICT tools to achieve them. One More framework was proposed by Abu-Shanab and Al-Dalou' that contained three levels and the appropriate technical tools necessary for every level. Furthermore, they recommended a catalog of performance statistics linked to every level of e-participation.

## Conclusions and Future Study

E-Government is about finding another useful way in which governments work together with the citizens, administrative agencies, companies, workers, and other participants. It enhances the democratic process and new proposals to create life simpler for people. The research area of e-Government is wide, extensive and small, and several scholars are engaged in various study projects in numerous issues in the area. The difficulty of e-participation practices findings from the multi-specialties included in this area. A general understanding was laid out about different electronic policy-making terminologies and their veracity of uses. Finally, more research is necessary for further awareness, knowledge, and understanding of such growing area, to concentrate on the significance of mobile involvement and social practices as two indispensables tools for the government to improve the involved political practices between the governments and the citizens. While defining the potential of democracy, governance and public work at the beginning of the information age is an amazing opportunity and responsibility. With the smart and efficient use of ICTs and in combination with democratic intent, we can create governments more approachable. We can also unite citizens to efficiently encounter public challenges, and eventually, we can build a more workable future for the advantage of the entire society and realm in which we live.

**AUTHOR'S PROFILE**

Sourav Mukherjee is a Senior Database Administrator and Data Architect based out of Chicago. He has more than 12 years of experience working with Microsoft SQL Server Database

Platform. His work focusses in Microsoft SQL Server started with SQL Server 2000. Being a consultant architect, he has worked with different Chicago based clients. He has helped many companies in designing and maintaining their high availability solutions, developing and designing appropriate security models and providing query tuning guidelines to improve the overall SQL Server health, performance and simplifying the automation needs. He is passionate about SQL Server Database and the related community and contributing to articles in different SQL Server Public sites and Forums helping the community members. He holds a bachelor's degree in Computer Science & Engineering followed by a master's degree in Project Management. Currently pursuing Ph.D. In Information Technology from the University of the Cumberlands. His areas of research interest include RDBMS, distributed database, Cloud Security, AI and Machine Learning. He is an MCT (Microsoft Certified Trainer) since 2017 and holds other premier certifications such as MCP, MCTS, MCDBA, MCITP, TOGAF, Prince2, Certified Scrum Master and ITIL.